# A thermodynamically consistent approach to describe the effect of thermal vacancy on abnormal thermodynamic behaviors of pure metals: application to body centered cubic W

Ying Tang [a,*], Lijun Zhang [b,*]

[a] School of Materials Science and Engineering, Hebei University of Technology, Tianjin 300130, P.R. China
[b] State Key Laboratory of Powder Metallurgy, Central South University, Changsha, Hunan 410083, P.R. China
* Corresponding authors
  E-mail: csuyingtang@163.com (Y. Tang); lijun.zhang@csu.edu.cn (L. Zhang)

**Abstract:**

In this paper, we developed a thermodynamically consistent approach to account for the Gibbs energy of pure metallic element with thermal vacancy over wide temperature range. Taking body centered cubic (bcc) W for a demonstration, the strong nonlinear increase for temperature dependence of heat capacities at high temperatures and a nonlinear Arrhenius plots of vacancy concentration in bcc W can be nicely reproduced by the obtained Gibbs energy. The successful description of thermal vacancy on abnormal thermodynamic behaviors in bcc W indicates that the presently proposed thermodynamically consistent approach is a universal one, and applicable to the other metals.

## 1. Introduction

Thermal vacancy is the simplest but extremely important structural defect in pure metals. As temperature increases, the thermal vacancy concentration in pure metals dramatically increases, and makes an apparent contribution to different physical quantities of materials, such as heat capacity, melting point, diffusivity, thermal conductivity, and so on [1-3]. Due to the important role of thermal vacancy in processes and properties of materials, it is attracting much attention recently [4-6]. Taking body centered cubic (bcc) W for instance, its thermal vacancy concentration can be larger than 0.02 at the melting point [7,8]. With such a large thermal vacancy concentration, the heat capacity of bcc W over the high-temperature region shows a strong non-linear increase [8-25], as demonstrated by the experimental data in Figure 1. As shown in the figure, the recent first-principles calculations [26] taking the harmonic/anharmonic vibration and electronic excitation into account cannot accurately predict the heat capacity of W with such abnormal non-linear increase over the high-temperature region. This fact indicates that the contribution from the thermal vacancy to heat capacity is non-negligible. In addition to heat capacity, the thermal vacancy also shows obvious influence on self-diffusivity of bcc W. The Arrhenius plot for measured self-diffusivities in bcc W over a wide temperature range shows significant curvature. Kraftmakher [3] pointed out that one probable reason for such curvature in self-diffusivities of bcc W lies in that the concentration of thermal vacancy has a similar temperature dependence.

Recent theoretical predications [27,28] show that formation entropy of vacancies is not constant as commonly assumed but increases with temperature, resulting in highly nonlinear temperature dependence in the formation energy, which naturally explains the dramatical increase of the heat capacity at high temperatures and curvature in Arrhenius plot of concentration vacancy. Consequently, in order to quantitatively describe the effect of thermal vacancy on these abnormal thermodynamic and diffusion behaviors in bcc W, accurate prediction of the temperature-dependent thermal vacancy formation energy is the prerequisite. Consequently, the major target of the present paper is to propose a thermodynamically-consistent approach to describe Gibbs energy and concentration of thermal vacancy, and their effects on abnormal behaviors in thermodynamic and diffusion properties of bcc W.

## 2. Model Description

The molar Gibbs free energy of pure element $k$ by considering the contribution of thermal vacancy is described as [4,29]:

$$G_m = \frac{1}{y_k}[y_k G_k + y_{va} G_{va} + RT(y_k \ln y_k + y_{va} \ln y_{va}) + y_k y_{va} \Omega] \quad (1)$$

where $R$ is the gas constant, $T$ is the absolute temperature, $y_k$ and $y_{va}$ are the site fractions of $k$ and thermal vacancy, respectively. The summation of $y_k$ and $y_{va}$ should be unity. $G_k$ is the molar Gibbs energy of the defect-free element $k$. While $G_{va}$ is the molar Gibbs energy of a virtual empty bcc lattice. In reality, it is very difficult to define the standard reference value for $G_{va}$ in a physical meaningful way [30]. In order to make the balance between the different terms in Eq. (1) at large temperature range, $G_{va}$ is suggested to be proportional to temperature[30]. Besides, it would be desirable to set a general value for $G_{va}$, which is independent of the elements. Recently, Franke [29] pointed out that the value of $G_{va}$ should be larger than a critical value (i.e., (ln2-1/2)$RT$) to ensure a unique equilibrium state. Such critical value is obtained from the pure mathematical analysis, and thus is a common one which does not depend on the types of elements or phases. In this work, the molar Gibbs energy of vacancy is simply set to be 0.2$RT$ [29], which is slightly larger than the critical value. $\Omega$ is the interaction parameter. If $\Omega$ is a constant, a constant mixing enthalpy is obtained. For the equilibrium state, one can have:

$$\frac{\partial G_m}{\partial y_{va}} = -\frac{1}{(1-y_{va})^2} G_{va} - \frac{1}{(1-y_{va})^2} RT \ln y_{va} - \Omega = 0 \quad (2)$$

Based on Eq. (2), the equilibrium concentration of thermal vacancy satisfies the following relation:

$$y_{va} = \exp(-\frac{G_{va} + \Omega(1-y_{va})^2}{RT}) \quad (3)$$

Meanwhile, the vacancy concentration can be also expressed by using the vacancy formation energy:

$$y_{va} = \exp(-G_{va}^f/RT) = \exp(-(H_f - TS_f)/RT) \quad (4)$$

where $G_{va}^f$ is the vacancy formation energy, $H_f$ and $S_f$ are the formation enthalpy and entropy of thermal vacancies, respectively. Comparing Eq. (3) with Eq. (4), it is easy to find that the value $G_{va} + \Omega$ approaches to the formation energy of thermal vacancy in the limit of negligible vacancy concentrations. However, it seems to be very difficult and even impossible to separate the contributions of $G_{va}$ and $\Omega$ from the formation energy of thermal vacancy[30]. The assumption of a constant entropy of formation of the thermal vacancy would introduce linear temperature dependence in $G_{va}^f$ and hence leads to a constant prefactor to the Arrhenius plot thermal vacancy concentration. However, for the Arrhenius plot of vacancy concentration with curvature behavior, i.e. $y_{va}$ in bcc W, such assumption cannot give a good description especially in the high temperature range. In order for an accurate thermodynamic description, the nonlinear temperature behavior of the vacancy formation energy needs to be considered. In a first approximation, we assume a linear temperature dependence for $S_f$, therefore, a temperature quadratic term will be included in $H_f$ as well as in $G_{va}^f$ due to the thermodynamic relation $(\partial H_f / \partial T)_P = T (\partial S_f / \partial T)_P$. As discussed above, we simply use the linear temperature dependence of $G_{va}$ in the present work. In order to describe the quadratic temperature behavior of vacancy formation energy, we thus propose a new expression for interaction parameter $\Omega$ in Eq. (1) :

$$\Omega = A + BT + CT^2 \quad (5)$$

Here, $A$, $B$ and $C$ are the parameters to be assessed on the basis of the experimental data, like thermal vacancy concentration.

Substituting Eq. (5) into Eq. (2), one can easily find that the proposed interaction expression is able to describe the curvature between $\log y_{va}$ and $1/T$. Then, based on the Gibbs energy expression (i.e., Eq. (1)) and the equilibrium state (i.e., Eq. (2)), the heat capacity of pure metal $k$ with thermal vacancy are derived as

$$C_P = C_p^{\text{Pure }k} - y_{va}T\frac{\partial^2\Omega}{\partial T^2} + (\Omega - T\frac{\partial\Omega}{\partial T})\frac{\partial y_{va}}{\partial T} \qquad (6)$$

in which $C_p^{\text{Pure }k}$ represents the heat capacity of defect-free element $k$. The second and third terms on the right hand side of Eq. (6) denote the effects of thermal vacancy on heat capacity. When submitting Eq.(5) to Eq.(6), it can be found that the effects of thermal vacancy on heat capacity will become obvious with the increase of the temperature, which is excepted to describe the abnormal behavior of thermodynamic properties near the melting point.

In the 1995's Ringberg Workshop [31], one heat capacity model for pure metals considering the physical contributions was proposed. In their model, the Debye or Einstein model are used to describe the heat capacity at low temperatures, and physically motivated polynomial is used to describe the high temperature data. It is well known that it is very difficult to integrate Debye model for Gibbs energy expression. In the present work, the above mentioned heat capacity model [31] with Einstein equation is thus used to describe $C_p^{\text{Pure }k}$:

$$C_p^{\text{Pure }k} = 3R\left(\frac{\theta_E}{T}\right)^2 \frac{e^{\theta_E/T}}{(e^{\theta_E/T}-1)^2} + aT + bT^2 + C_p^{magn} \qquad (7)$$

The first term in Eq.(7) is the Einstein heat capacity, which is mostly contributed from harmonic vibration, and $\theta_E$ is the Einstein temperature. The second term $aT$ in Eq.(7) is related to electronic excitations and low order anharmonic corrections, while the third term $bT^2$ contains the next order anharmonic corrections. $C_p^{magn}$ is the contribution from the magnetic ordering. It should be noted that, in order to obtain the defect-free heat capacity expression, the coefficients $a$ and $b$ need to be evaluated by considering the experiemntal data without any contribution from thermal vacancy.

Then, the Gibbs energy for pure element can be evaluated from the expression of heat capacity:

$$G_k = E_0 + \frac{3}{2}R\theta_E + 3RT\ln[1-\exp(-\frac{\theta_E}{T})] - \frac{a}{2}T^2 - \frac{b}{6}T^3 - \int_0^T[\int_0^T \frac{C_p^{mag}}{T}dT]dT \qquad (8)$$

where $E_0$ is the total energy of ferromagnetic pure $k$ at 0 K which can be directly obtained from the first principles calculations, while the second term is the energy of zero-point lattice vibration [32, 33].

### 3. Application to bcc W

The above proposed thermodynamic description for pure metallic element with thermal vacancy is then applied to bcc W as an example. The heat capacity of bcc W was experimentally measured by several groups [8-25] from 0 K to melting points, as shown in Figure 1. The heat capacity shows a rapid increase at high temperatures especially close to the melting point. Besides, the enthalpy of bcc W was also measured over wide temperature range[15, 34-39]. Moreover, the thermal vacancy concentration in bcc W was also experimentally investigated at melting temperature [7,8]. Though no experimental vacancy concentration as a function of temperature was reported in the literature, Kraftmakher [3] derived the equilibrium vacancy concentrations in bcc W from the heat capacities with nonlinear increase [8].

Theoretically, the heat capacity and Gibbs energy of defect-free bcc W should exclude any contribution from thermal vacancy. In reality, it is also very difficult to separate the contribution of thermal vacancy in the experimental data completely. The purely first-principles calculations may provide the data for defect-free bcc W, but the current first-principles computed heat capacities of defect-free bcc W available in the literature [26] are lower than the experimental data (see Figure 1) above 1500 K. As pointed out by Kraftmakher[3], the vacancy contribution to specific heat becomes visible only at temperatures above about two third of the melting temperature (i.e., 2463 K for bcc W). Besides, the magnetic effects on heat capacity can be ignored for bcc W. Moreover,

Walford[40] experimental measured and reported that the Debye temperature ($\theta_D$) of pure bcc W was 377 K. Then, the Einstein temperature ($\theta_E$) in Eq.(7) and (8) is set to be 269.2 K by using the relation $\theta_E \approx 0.714\theta_D$ [41]. Subsequently, the heat capacity of defect-free bcc W was obtained by fitting the experimental heat capacities and enthalpies below 2/3 $T_m$. There is no measured total energy ($E_0$ in Eq.(8)) of bcc W at 0 K in the literature. Wang et al.[42] reported the total energy of bcc W at 0 K by using the first-principle calculation, which is used in the present work to get the expression for Gibbs energy of defect-free bcc W. After that, the three coefficients constituting the interaction parameter, $\Omega$, were then evaluated based on the experimental heat capacities over the range of 2/3$T_m$~$T_m$. The finally obtained thermodynamic parameters in Gibbs energy expression for bcc W are listed Table 1. The coefficient of $C$ in Eq. (4) was introduced to describe the strong nonlinear behavior of heat capacity at high temperatures.

Figure 1 shows the calculated heat capacity for bcc W using the presently obtained Gibbs energy expression in comparison with the experimental data [8-25]. Excellent agreement between the calculation and the experiments is obtained. For a comparison, the calculated heat capacity of defect-free bcc W is also superimposed in Figure 1. The deviation between the heat capacity with and without thermal vacancy is quite obvious at high temperatures, and can reach to 26.5 J (mol K)$^{-1}$ at melting temperature. It indicates that the thermal vacancy has a significant effect on the thermodynamic properties of the pure metals. Figure 2 shows the presently calculated heat contents (H-H$_{298}$) of bcc W with and without thermal vacancy along with experimental data [15, 34-39]. As shown in Figure 2, the calculated results with thermal vacancy reproduce excellent with the reported data, while the results without thermal vacancy show a small deviation from the experimental data at the high temperature range. It further indicates the reasonability of the present proposed model. Figure 3 shows the presently predicted 10-base logarithm values of thermal vacancy concentration of bcc W as a function of 10000/T, compared with the limited experimental data at melting temperature [7,8]. As can be seen in the figure, the calculated thermal vacancy concentration of bcc W is 0.018, which agrees reasonably with the experimental data [7,8]. Moreover, as expected, such Arrhenius plot of vacancy concentration shows a clear curvature.

## 4. Conclusion

In summary, a thermodynamic-consistent approach was proposed in this paper to describe the Gibbs energy of pure metallic element with thermal vacancy over a wide temperature range. The proposed approach was then applied to obtain the new thermodynamic descriptions for bcc W, which can well reproduce the curvature for Arrhenius plots of the vacancy concentrations, as well as the strong nonlinear temperature-dependence thermodynamic properties such as heat capacity close to the melting point. It should be noted that the presently proposed thermodynamically consistent approach to describe the effect of thermal vacancy on Gibbs energy is a universal one, and also applicable to all the other pure metals.

**Acknowledgements**

The financial support from the National Key Research and Development Program of China (Grant No. 2016YFB0301101) and the Hunan Provincial Science and Technology Program of China (Grant No. 2017RS3002) - Huxiang Youth Talent Plan is acknowledged. Y. Tang acknowledges the financial support from Yuanguang fellowship released by Hebei University of Technology. L. Zhang acknowledges the project supported by State Key Laboratory of Powder Metallurgy Foundation, Central South University, Changsha, China.

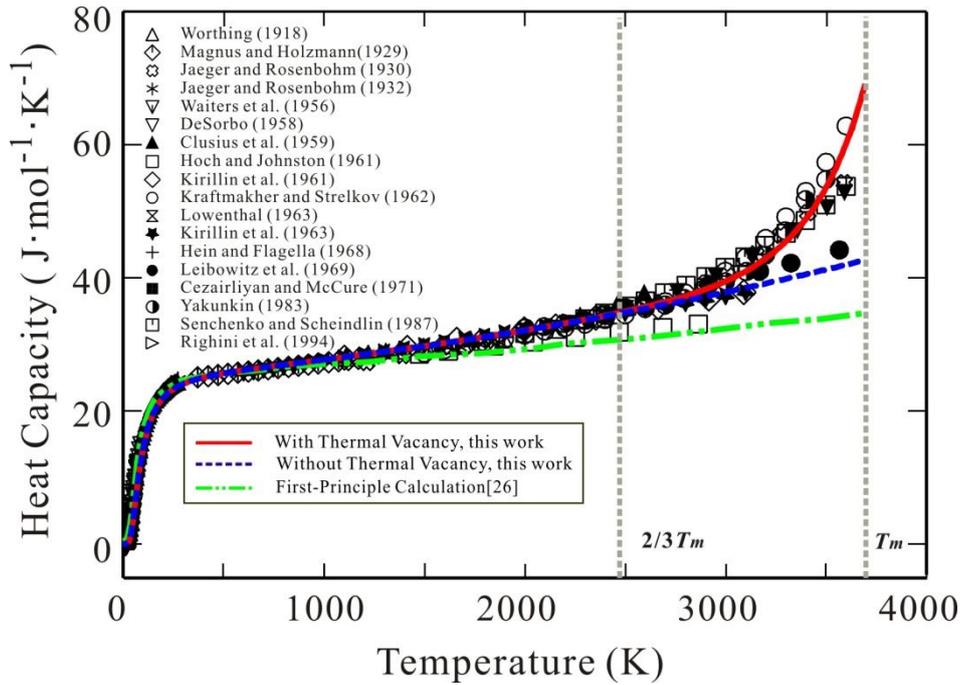

**Figure 1.** Heat capacity of bcc W as a function of temperature. Symbols: Experimental data[8-25]. Solid line (red): Calculated results according to the presently established Gibbs energy for bcc W with thermal vacancy contribution; Dashed line (blue): Calculated results according to the presently established Gibbs energy for defect-free bcc W without thermal vacancy contribution; Dotted line (green): First-principles calculation[26], which does not include thermal vacancy contribution.

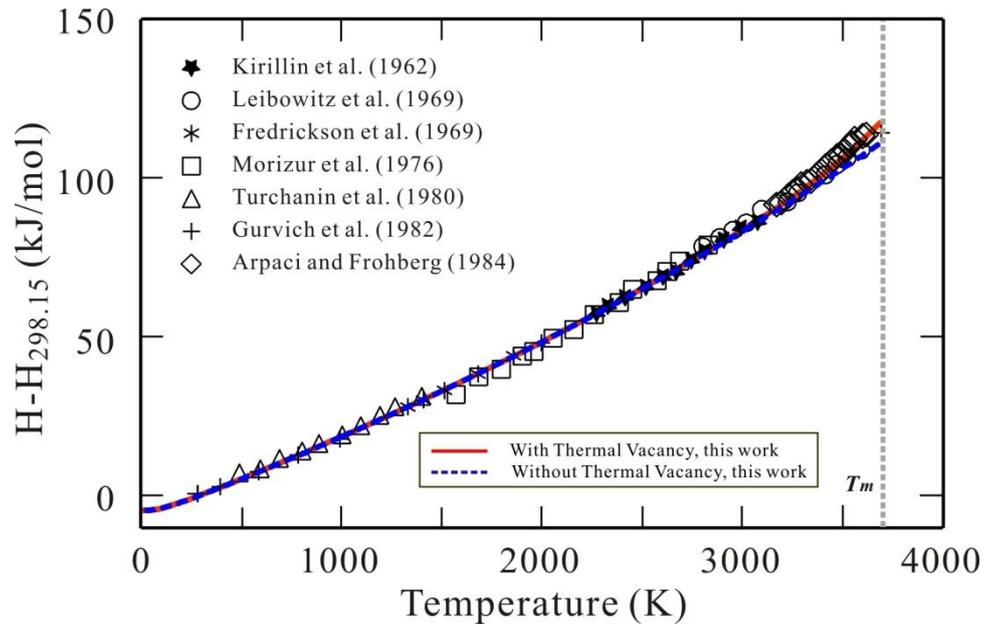

**Figure 2.** Heat contents (H-H$_{298.15}$) of bcc W as a function of temperature. Symbols: Experimental data[15,34-39]. Solid line (red): Calculated results according to the presently established Gibbs energy for bcc W with thermal vacancy contribution; Dashed line (blue): Calculated results according to the presently established Gibbs energy for defect-free bcc W without thermal vacancy contribution.

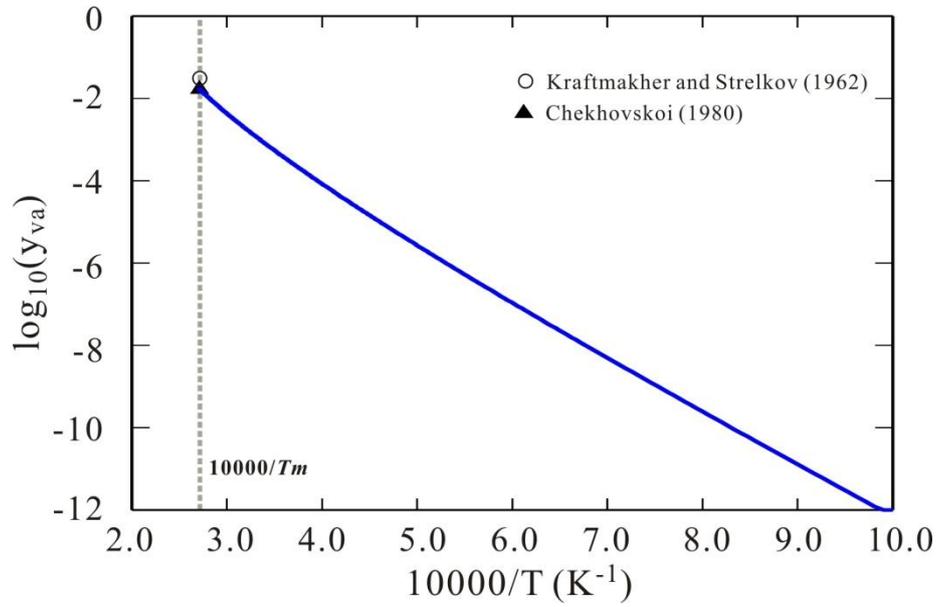

**Figure 3.** Arrhenius plots of thermal vacancy concentration of bcc W. Symbols: Experimental data at melting point from[7,8]. Solid line: Calculated thermal vacancy concentration of bcc W according to the presently established Gibbs energy for bcc W. A clear curvature is obtained for the Arrhenius plot of thermal vacancy concentration of bcc W.

**Table 1.** List of the evaluated thermodynamic parameters for bcc W.

| Parameters | Values (Gibbs energy in J/mol-atom; *T* in Kelvin) |
|---|---|
| $G_W$ | $E_0 + \frac{3}{2}R\theta_E + 3RT\ln(1-\exp(-\frac{\theta_E}{T})) - 1.085\times10^{-3}T^2 - 1.1835\times10^{-7}T^3$ <br> ($E_0 = -1228665.43, \quad \theta_E = 269.2$) |
| $G_{Va}$ | $+0.2RT$ |
| $\Omega$ | $+229615.89 + 12.73T - 1.1274\times10^{-2}T^2$ |